\documentclass[prl,aps,twocolumn,superscriptaddress]{revtex4}
\usepackage{}

\usepackage{graphicx}  
\usepackage{epstopdf}
\usepackage{dcolumn}   
\usepackage{bm}        
\usepackage{amssymb}   
\usepackage{amsmath}   
\usepackage{amsthm}
\usepackage{chemarrow}
\usepackage{color}
\usepackage{mathrsfs}
\usepackage{float}
\usepackage{bbold}
\usepackage{bbm}
\usepackage[a4paper,colorlinks=true,
linkcolor=blue,citecolor=blue,
pdfauthor={ },
pdftitle={ },
pdfsubject={ },
pdfkeywords={ }]{hyperref}

\theoremstyle{plain}
\newtheorem*{theorem*}{Theorem}

\begin{document}


\title{Ultralow-field Nuclear Magnetic Resonance Asymmetric Spectroscopy}


\author{Min Jiang}
\email[]{These authors contributed equally to this work}
\affiliation{
Hefei National Laboratory for Physical Sciences at the Microscale and Department of Modern Physics, University of Science and Technology of China, Hefei 230026, China}
\affiliation{
CAS Key Laboratory of Microscale Magnetic Resonance, University of Science and Technology of China, Hefei, Anhui 230026, China}

\author{Wenjie Xu}
\email[]{These authors contributed equally to this work}
\affiliation{
Hefei National Laboratory for Physical Sciences at the Microscale and Department of Modern Physics, University of Science and Technology of China, Hefei 230026, China}
\affiliation{
CAS Key Laboratory of Microscale Magnetic Resonance, University of Science and Technology of China, Hefei, Anhui 230026, China}

\author{Yunlan Ji}
\affiliation{
Hefei National Laboratory for Physical Sciences at the Microscale and Department of Modern Physics, University of Science and Technology of China, Hefei 230026, China}
\affiliation{
CAS Key Laboratory of Microscale Magnetic Resonance, University of Science and Technology of China, Hefei, Anhui 230026, China}

\author{Ji Bian}
\affiliation{
Hefei National Laboratory for Physical Sciences at the Microscale and Department of Modern Physics, University of Science and Technology of China, Hefei 230026, China}
\affiliation{
CAS Key Laboratory of Microscale Magnetic Resonance, University of Science and Technology of China, Hefei, Anhui 230026, China}

\author{Shiming Song}
\affiliation{
Hefei National Laboratory for Physical Sciences at the Microscale and Department of Modern Physics, University of Science and Technology of China, Hefei 230026, China}
\affiliation{
CAS Key Laboratory of Microscale Magnetic Resonance, University of Science and Technology of China, Hefei, Anhui 230026, China}

\author{Xinhua Peng}
\email[]{xhpeng@ustc.edu.cn}
\affiliation{
Hefei National Laboratory for Physical Sciences at the Microscale and Department of Modern Physics, University of Science and Technology of China, Hefei 230026, China}
\affiliation{
CAS Key Laboratory of Microscale Magnetic Resonance, University of Science and Technology of China, Hefei, Anhui 230026, China}
\affiliation{
Synergetic Innovation Center of Quantum Information and Quantum Physics, University of Science and Technology of China, Hefei, Anhui 230026, China}
\affiliation{
Synergetic Innovation Center for Quantum Effects and Applications, Hunan Normal University, Changsha 410081, China}

\begin{abstract}
Ultralow-field nuclear magnetic resonance (NMR) provides a new regime for many applications ranging from materials science to fundamental physics.
However, the experimentally observed spectra show asymmetric amplitudes, differing greatly from those predicted by the standard theory.
Its physical origin remains unclear, as well as how to suppress it.
Here we provide a comprehensive model to explain the asymmetric spectral amplitudes,
further observe more unprecedented asymmetric spectroscopy and find a way to eliminate it.
Moreover, contrary to the traditional idea that asymmetric phenomena were considered as a nuisance,
we show that more information can be gained from the asymmetric spectroscopy, e.g., the light shift of atomic vapors and the sign of Land$\acute{\textrm{e}}$ $g$ factor of NMR systems.
\end{abstract}


\maketitle

\textsl{Introduction.\textbf{--}}Nuclear magnetic resonance (NMR) is a fundamental exploratory tool in detecting, identifying, and quantifying information about the atoms and molecules~\cite{Ernst1987}.
With the advent of hyperpolarization methods~\cite{Adams2009, Shchepin2015, Theis2012}, detection schemes using atomic magnetometers~\cite{Budker2007, Allred2002, Kominis2003, Shah2007} or superconducting quantum interference devices (SQUIDs)~\cite{Greenberg1998} and quantum control techniques~\cite{Sjolander2017, Tayler2016, Bian2017, Jiang2017, Jiang2018, Ji2018},
ultralow-field NMR has been developed as an alternative magnetic resonance modality~\cite{McDermott2002, Burghoff2005, Appelt2006, Theis2011, Ledbetter2011, BlanchardD2016, Bevilacqua2009}.
Atomic magnetometers are an ideal tool because, in contrast to SQUIDs, they do not require cryogenically cooling.
Recent works using atomic magnetometers in ultralow-field $\textrm{NMR}$ spectroscopy have been extensively reported in Refs.~\cite{Theis2011, Ledbetter2011, Ledbetter2009, Liu2013}.
Ultralow-field $\textrm{NMR}$ experiments regularly achieve nuclear spin coherence times longer than ten seconds~\cite{Blanchard2013, Emondts2014},
and thus can be used for chemical fingerprinting and precise measurement of nuclear spin-spin couplings~\cite{Theis2011, Blanchard2013, Blanchard2015}.
Very recently, ultralow-field NMR has attracted renewed interest for application in fundamental physics,
such as searches for molecular chirality~\cite{King12017}, ultralight axions and axion-like particles~\cite{Garcon2017} and nuclear spin-gravity coupling~\cite{Teng2018}.

Standard ultralow-field $\textrm{NMR}$ theory has been developed to analyse the spectroscopy~\cite{Ledbetter2011, BlanchardD2016, Appelt2010},
i.e., predict the frequencies and amplitudes of resonant peaks.
However, researchers are disconcerted to find that ultralow-field NMR spectra of even a sample as simple as formic acid (containing $^{13}$C-$^1$H spin pairs)£¬
experimentally suffer from asymmetric amplitudes,
differing greatly from those predicted by the standard theory~\cite{Teng2018, Ledbetter2012, Jiang22018}.
Thus, there is an urgent need to understand the asymmetric amplitude phenomena in ultralow-field NMR spectra.

In this Letter, we provide a comprehensive model for interpreting the asymmetric effect by investigating the asymmetric amplitude phenomena in ultralow-field $\textrm{NMR}$ spectroscopy,
where a class of unprecedented asymmetric phenomena are also present.
We further find that the asymmetric phenomena can be completely eliminated when the external magnetic field is carefully chosen,
as observed experimentally.
Moreover, the asymmetric spectroscopy can surprisingly impart additional information,
including the light shift of atomic vapors and the sign of Land$\acute{\textrm{e}}$ $g$ factor of NMR systems.

\begin{figure}[t]  
	\makeatletter
	\def\@captype{figure}
	\makeatother
	\includegraphics[scale=0.62]{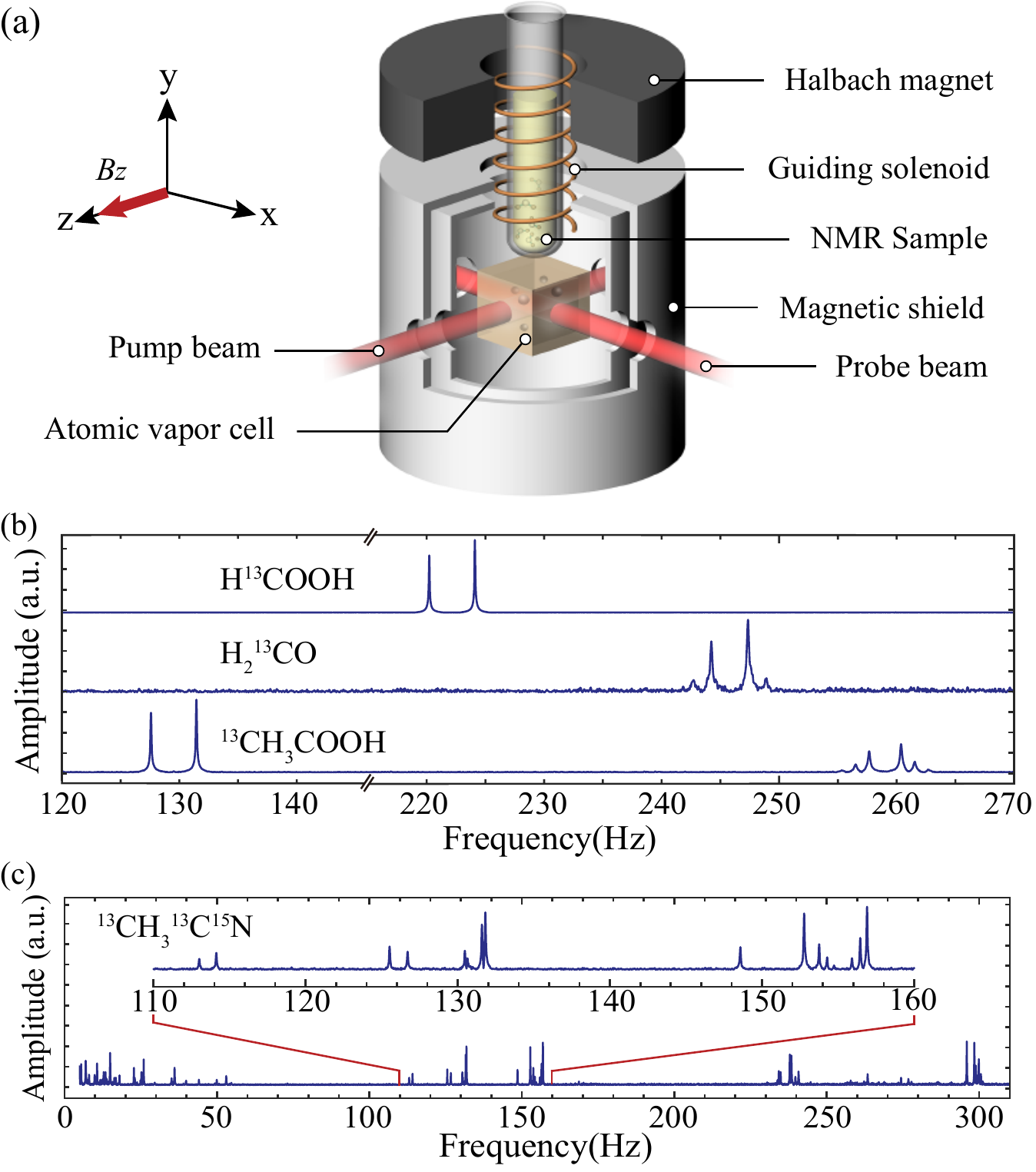} %
	\caption{(color online). (a) Diagram of the ultralow-field $\textrm{NMR}$ spectrometer. After shuttling, the NMR sample is located at a distance of 1~mm above a rubidium ($^{87}$Rb) vapor cell, which resistively heated to ${190^ \circ }\textrm{C}$. The $^{87}$Rb atoms are pumped by $z$-directed, circularly polarized laser light at the D1 transition. The magnetic field is measured via optical rotation of linearly polarized probe laser light at the D2 transition. (b) Asymmetric ultralow-field NMR spectra of formic acid (H$^{13}$COOH), formaldehyde (H$_2$$^{13}$CO), acetic acid ($^{13}$CH$_3$COOH), as examples of $^{13}$CH$_n$ systems. (c) Asymmetric ultralow-field NMR spectrum of fully labeled acetonitrile ($^{13}$CH$_3$$^{13}$C$^{15}$N). A zoom shows the central part of the spectrum. }
	\label{fig1}
\end{figure}

\textsl{Asymmetric spectra in ultralow-field NMR.\textbf{--}}A liquid-state $n$-spin system at a magnetic field can be described by the Hamiltonian
\begin{equation}
H = \sum\limits_{i;j > i} {2\pi {J_{ij}}} {\textbf{I}_i} \cdot {\textbf{I}_j} - \sum\limits_j {{\gamma _j}{\textbf{I}_j} \cdot \textbf{B}},
\label{H}
\end{equation}
where ${J_{ij}}$ is the strength of the scalar spin-spin coupling ($J$ coupling) between the $i$th and $j$th spins,
$\mathbf{I}_j=(I_{jx}, I_{jy}, I_{jz})$ represent the $j$th spin with gyromagnetic ratio $\gamma _j$,
and the reduced Planck constant is set to one.
At zero magnetic field,
eigenstates are also eigenstates of $\textbf{f}^2$ and $f_z$,
where $\textbf{f}$ are the total angular momentum,
and are denoted as $\left| {f,{m_f}} \right\rangle$.
In the presence of a perturbing magnetic field, eigenstates are approximately those of the Hamiltonian described in Eq.~(\ref{H}) at zero field,
and energies can be calculated with degenerate perturbation theory~\cite{Ledbetter2011, Appelt2010}.
The ultralow-field NMR spectrometer used in our experiments is similar to the apparatus in Refs.~\cite{Ledbetter2011, TaylerMC2017} and is depicted in Fig.~\ref{fig1}(a).
Liquid-state NMR samples are contained in 5-mm NMR tubes,
and pneumatically shuttled between a prepolarizing magnet ($B_p\approx 1.3$~T) and a rubidium ($^{87}$Rb) vapor cell.
During the transfer, a guiding magnetic field ($\sim 1$~G) is applied along the $y$ axis,
and is abruptly switched off within 10~$\mu$s prior to signal acquisition.
In the high-temperature approximation, the initial spin state is $\rho_0=1/2^n(\mathbbm{1}+\sum_j \epsilon_j I_{jy})$ with $\epsilon_j=\gamma_j B_p/k_B T \sim 10^{-6}$,
where $k_B$ is the Boltzmann constant, and $T$ is the temperature of the sample.
The initial spin state then evolves under the Hamiltonian in Eq.~(\ref{H}),
and generates a magnetization with component ${M_\zeta }(t)$ along $\zeta$ axis ($\zeta=x, y, z$).
The magnetization is detected with a spin-exchange relaxation-free atomic magnetometer~\cite{Allred2002, Kominis2003} (sensitivity $\approx$ $25~\textrm{fT}/\sqrt{\textrm{Hz}}$),
and Fourier transformed to the NMR spectrum of the sample.

\begin{figure}[t]  
	\makeatletter
	\def\@captype{figure}
	\makeatother
	\includegraphics[scale=0.66]{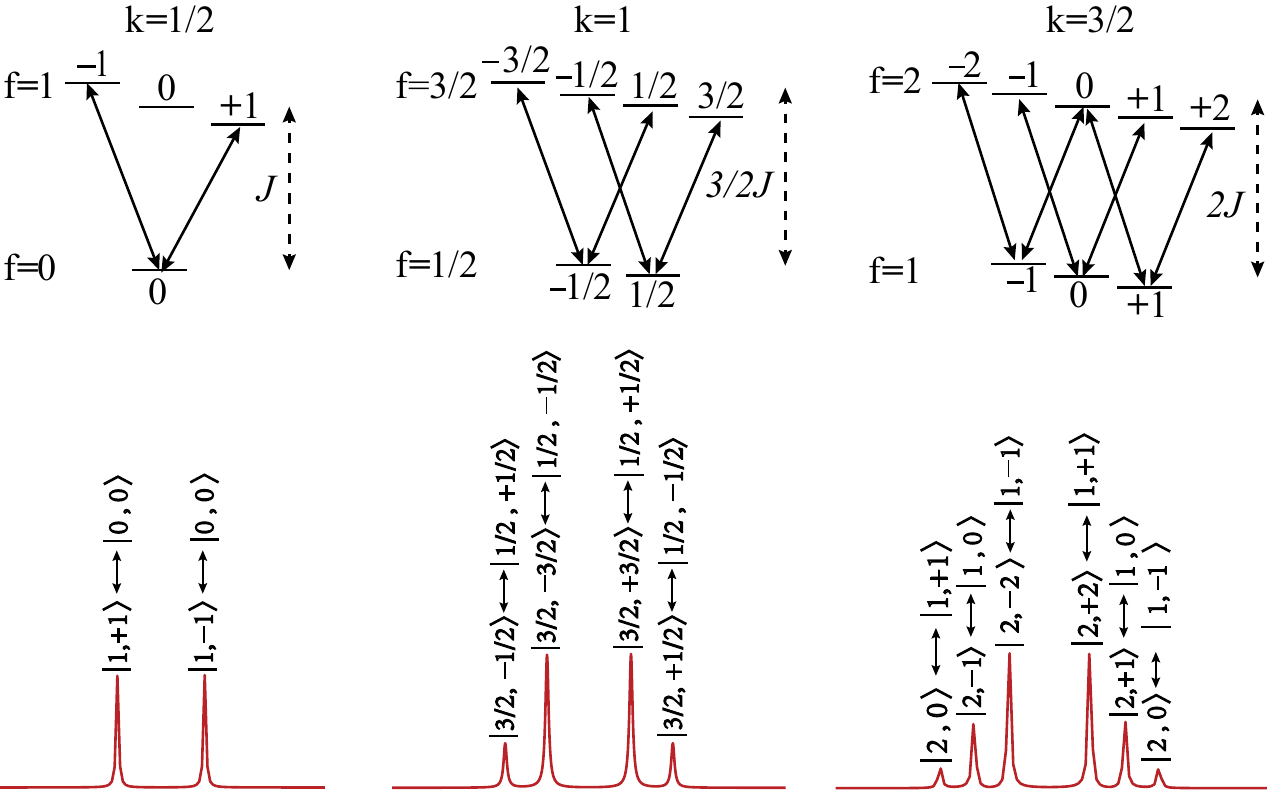} %
	\caption{(color online). Energy levels for $^{13}$CH$_n$ systems and the corresponding spectra predicted from the standard theory (see also Refs.~\cite{Ledbetter2011, Appelt2010}). Quantum number $k$ is the total angular momentum of $n$ equivalent proton spins. The system contains a manifold with $k=1/2$ for $n =1$, a manifold with $k=1$ for $n=2$, and two manifolds with $k=1/2$ and $k=3/2$ for $n=3$.  }
	\label{fig2}
\end{figure}

We experimentally investigate the ultralow-field spectra of typical $^{13}\textrm{C}{\textrm{H}_n}$ systems.
In such systems, $n$ equivalent proton spins couple to a carbon spin with the same strength $J$.
Figure~\ref{fig1}(b) shows the experimental spectra of formic acid, a doublet centered at $J\approx 222.2$~Hz,
where the characteristic of asymmetric amplitudes is obvious.
Taking acetic acid as a $^{13}\textrm{C}{\textrm{H}_3}$ example to illustrate,
the corresponding spectra of the $k=1/2$ manifold, centered at $J\approx 129.5$~Hz, is same with the case of formic acid.
We focus on the $k=3/2$ manifold, in which the six lines centered at $2J\approx 259.0$~Hz can be divided into three pairs,
and each pair consists of two transitions of $\left| {2,{m_f}} \right\rangle  \leftrightarrow$ $ \left| {1,m_f+1} \right\rangle $ and $\left| {2, - {m_f}} \right\rangle  \leftrightarrow $$\left| {1, - m_f-1} \right\rangle $,
where $m_f=-2,-1,0$.
These three pairs of NMR lines show asymmetric amplitudes.
The asymmetric phenomena are not limited to the $^{13}$CH$_n$ systems.
Figure~\ref{fig1}(c) shows the asymmetric spectra experimentally measured in a more complex spin system (fully labeled acetonitrile) which demonstrates the asymmetric phenomena is ubiquitous for ultralow-field NMR in this kind of setup.
However, for these systems, the $\textrm{NMR}$ spectral asymmetry is significantly inconsistent with the theoretical predictions, as shown in Fig.~\ref{fig2}.
Even considering high-order corrections to the eigenstates which introduce some asymmetry,
it still cannot account for the experimental data.

\begin{figure}[t]  
	\makeatletter
	\def\@captype{figure}
	\makeatother
	\includegraphics[scale=0.85]{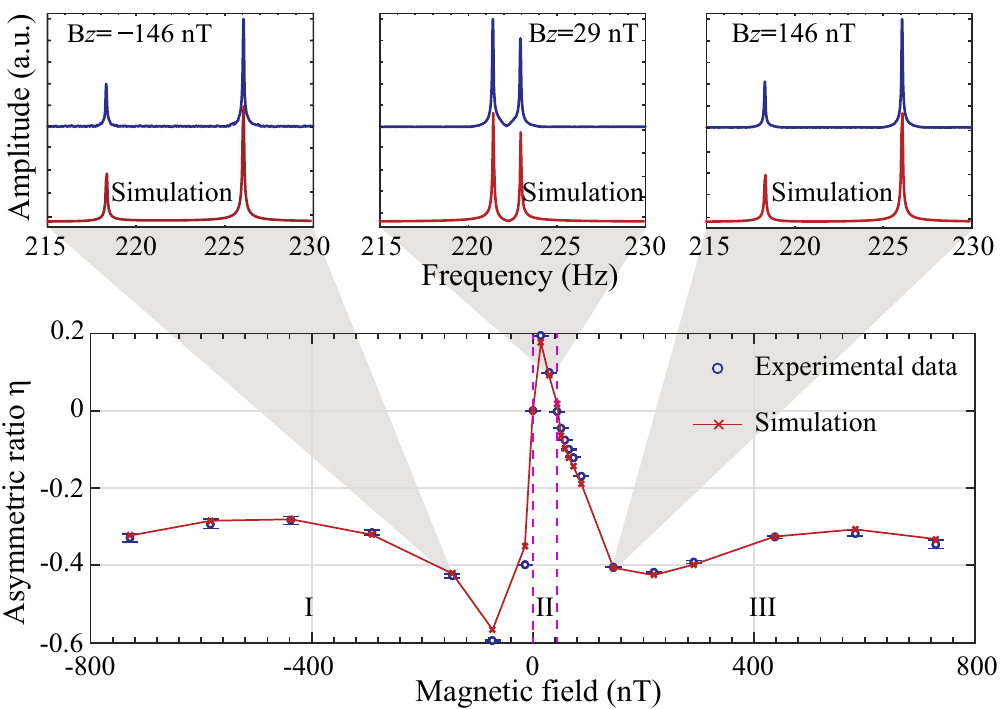} %
	\caption{(color online). Plot of the asymmetric ratios of formic acid doublet as a function of magnetic fields. The experimental data are shown with blue points, which are in good agrement with the calculations (red line) based on our comprehensive model. The insets are the ultralow-field NMR spectra at corresponding magnetic fields.}
	\label{fig3}
\end{figure}

To investigate the origin of the asymmetric phenomena,
we restrict our attention here to formic acid as the simplest example of general phenomena.
We define an amplitude ratio of the doublet,
$\eta=({\textit{Amp}_1} - {\textit{Amp}_2})/({\textit{Amp}_1} + {\textit{Amp}_2})$,
as a metric of the asymmetry.
Here, ${\textit{Amp}_1}$ and $\textit{Amp}_2$ are the amplitudes of the peaks at lower frequency and higher frequency, respectively.
The asymmetric ratios $\eta$ are plotted as a function of magnetic fields, as shown with the blue circles in Fig.~\ref{fig3}.
We find that the asymmetry of the doublet shows strongly dependence on the applied magnetic field.
For example,
in the left inset of Fig.~\ref{fig3}, the amplitude of the peak at lower frequency is smaller than that of the peak at higher frequency when $B_z=-146~\textrm{nT}$.
We call this phenomena as negative asymmetry (i.e., $\eta<0$), and on the contrary we call it as positive asymmetry (i.e., $\eta>0$).
When $B_z=29~\textrm{nT}$, the doublet exhibit positive asymmetry in the middle inset of Fig.~\ref{fig3}.
As described in the Supplemental Material~\cite{SI},
we also examine the asymmetry in acetic acid and the experimental results are similar with the case of formic acid.

We now make three observations on the plot of $\eta$ as a function of magnetic fields.
(1) The plot has two cross points  (e.g., $B_z=0$ and $B_z \approx 43.7~\textrm{nT}$ in Fig.~\ref{fig3}) with $\eta=0$, in which the corresponding spectra are symmetric.
At the zero-field cross point, the spectrum corresponds to the zero-field NMR,
which is trivially symmetric for only a single peak is observed.
The non-zero cross point corresponds to a specific magnetic field, which allows for symmetric ultralow-field spectra.
Further discussions are presented in below.
(2) The plot is divided into three regions by the two cross points.
The spectral doublet have same asymmetry in the regions I and III, and have opposite asymmetry in the region II.
(3) The ultralow-field spectra of samples (e.g., formic acid, formaldehyde) tend to exhibit negative asymmetry in the regions I and III.
We now present a comprehensive model (with a critical consideration of the magnetometer's frequency response) to interpret the above-mentioned asymmetric phenomena.

\textsl{A comprehensive model for explaining asymmetric spectroscopy.\textbf{--}}To explain the asymmetric $\textrm{NMR}$ spectra,
we firstly consider the effect of the amplitude-frequency and phase-frequency responses of the atomic magnetometer~\cite{Allred2002,Jiang22018, SI, Li2006}.
Recently, the phase-frequency response was used to correct the phase of zero-field NMR spectra~\cite{Theis2011},
and compensate the phase difference between two magnetometer channels~\cite{Jiang22018}.
Application of a magnetic field ${B_z}$ makes the atomic magnetometer simultaneously sensitive to fields along the $x$ and $y$ axes~\cite{SI}.
This has been applied to realize a vector atomic magnetometer~\cite{Seltzer2004}.
Here, we demonstrate that the phase-frequency response is different along the $x$ and $y$ axes depending on the magnetic field ${B_z}$.
This point is ignored until we recognize that it is crucial for explaining the asymmetric NMR spectra (see below).
We denote the amplitude-frequency response along $\zeta$ axis as $A_\zeta(\nu ,{B_z})$ and the phase-frequency response as $\Phi_\zeta(\nu ,{B_z})$.
In the following discussion, a more careful analysis shows that $A_\zeta(\nu ,{B_z})$ and $\Phi_\zeta(\nu ,{B_z})$ are key to explaining experimental asymmetric spectra.

We then calculate the magnetization signals generated by the NMR system.
Calculating the spin magnetization evolution under the Hamiltonian as described in Eq.~(\ref{H}) shows that the formic acid simultaneously generate ${M_x ^{{\nu _i}}}(t)$ and ${M_y ^{{\nu _i}}}(t)$ $(i=1,2)$ with same amplitude.
Here,
${M_{x(y)} ^{{\nu _i}}}(t)$ is the component of $M_{x(y)}(t)$ with the frequency $\nu_i$,
$\nu_1$ and $\nu_2$ correspond to transition  $\left| {1, 1} \right\rangle  \leftrightarrow \left| {0,0} \right\rangle $ and $\left| {1,-1} \right\rangle  \leftrightarrow \left| {0,0} \right\rangle $, respectively.
The initial phase of ${M_x ^{{\nu _1}}}(t)$ is $\pi/2$ ahead of the ${M_y ^{{\nu _1}}}(t)$,
while the initial phase of ${M_x ^{{\nu _2}}}(t)$ is $\pi/2$ behind of the ${M_y ^{{\nu _2}}}(t)$ (see Supplemental Material~\cite{SI}).
${M_x ^{{\nu _i}}}(t)$ and ${M_y ^{{\nu _i}}}(t)$ generate magnetic fields on the vapor cell along $x$ axis, ${B_x ^{{\nu _i}}}(t)$, and $y$ axis, ${B_y ^{{\nu _i}}}(t)$, respectively.
The magnetic field along $y$ is different from the magnetic field along $x$,
i.e., ${|{B_x ^{{\nu _i}}}(t)|} = \kappa{|{B_y ^{{\nu _i}}}(t)|}$.
Here, $\kappa$ depends on the spatial configuration of the sample and the vapor cell.
Therefore, the oscillating magnetic field produced by the spin magnetization on the vapor cell can be written as
\begin{equation*}
\left[ {\begin{array}{*{20}{c}}
   {B_x^{{\nu _i}}(t)}  \\
   {B_y^{{\nu _i}}(t)}  \\
\end{array}} \right] \propto \left[ {\begin{array}{*{20}{c}}
   1 & 0  \\
   0 & \kappa   \\
\end{array}} \right] \cdot \left[ {\begin{array}{*{20}{c}}
   {M_x^{{\nu _i}}(t)}  \\
   {M_y^{{\nu _i}}(t)}  \\
\end{array}} \right] \propto \left[ {\begin{array}{*{20}{c}}
   {{e^{ - i2\pi {\nu _i}t + i{\theta _i}}}}  \\
   {\kappa {e^{ - i2\pi {\nu _i}t}}} \\
\end{array}} \right]+\textrm{c.c.},
\end{equation*}
where
$\theta _1= \pi/2$, and
$\theta _2= -\pi/2$.

With taking the amplitude-frequency response $A_\zeta(\nu ,{B_z})$ and phase-frequency response $\Phi_\zeta(\nu ,{B_z})$ of the atomic magnetometer,
the magnetometer signal is proportional to
\begin{equation}
\sum\limits_i {{{\bf{P}}_i}^T\cdot\left[ {\begin{array}{*{20}{c}}
   1 & 0  \\
   0 & \kappa   \\
\end{array}} \right]\cdot\left[ {\begin{array}{*{20}{c}}
   {{e^{ - i2\pi {\nu _i}t + i{\theta _i}}}}  \\
   {{e^{ - i2\pi {\nu _i}t}}}  \\
\end{array}} \right]}  + \textrm{c.c.},
\label{sig}
\end{equation}
where
\begin{equation*}
\textbf{P}_i = \left[ {\begin{array}{*{20}{c}}
   {{A_x}(\nu_i ,{B_z}){e^{ - i{\Phi_x}(\nu_i ,{B_z})}}}  \\
   {{A_y}(\nu_i ,{B_z}){e^{ - i{\Phi_y}(\nu_i ,{B_z})}}}  \\
\end{array}} \right].
\end{equation*}
Therefore, each NMR peak of the formic acid doublet at the frequency $\nu_i$ is the interference between the magnetometer's response of the spin magnetization along $x$ and $y$ axes.
As such, the interference effect causes the asymmetry of the doublet peaks of formic acid.
Moreover, $A_\zeta(\nu ,{B_z})$ and $\Phi_\zeta(\nu ,{B_z})$ vary with the magnitude of the external magnetic field ${B_z}$,
and thus the asymmetric ratio of doublet peaks varies with the magnitude of the external magnetic field ${B_z}$.
Using the experimental data~\cite{SI} of $A_\zeta(\nu ,{B_z})$, $\Phi_\zeta(\nu ,{B_z})$ and $\kappa$,
we calculate the spectral asymmetric ratios with respect to the magnetic field,
as shown in Fig.~\ref{fig3}.
The experimental results are in good agreement with the theoretical calculations by Eq.~(\ref{sig}).
Therefore, from these experimental results, our model clearly provides the explanation of the origin of the asymmetric amplitude phenomena in ultralow-field NMR spectroscopy.

The model presented above can be generalized to general quantum sensors.
Ordinarily, the standard process of quantum mechanics treats a quantum sensor as an observable operator,
and then focus on the dynamics of the detected quantum system.
However, quantum sensors, such as $\textrm{SQUIDs}$~\cite{Greenberg1998} and nitrogen-vacancy centres~\cite{Taylor2008},
have frequency responses to the oscillating signals generated by detected systems~\cite{Budker2007, Degen2017}.
The normal observable operator has no ability to include the effect of the frequency response of quantum sensors.
Here we show that the effect of frequency response is effectively equal to modifying the observable operator by applying specific operations.
For example, the normal observable operator is the spin angular moment $I_y$, which corresponds to the observation of $y$ magnetization.
After introducing the phase-frequency response,
the effective observable operator is the one after
rotating $I_y$ with $\Phi_y(\nu ,{B_z})$ angle around $z$ axis~\cite{SI}.

\begin{figure}[t]  
	\makeatletter
	\def\@captype{figure}
	\makeatother
	\includegraphics[scale=1.03]{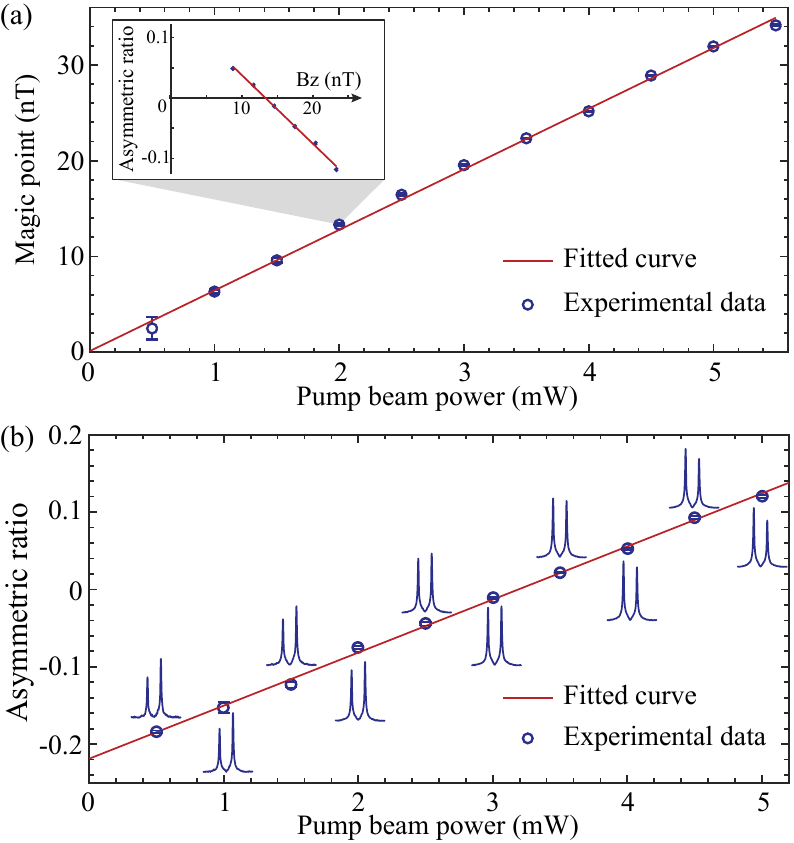} %
	\caption{(color online). (a) The experimental asymmetric ratios for formic acid doublet with respect to different pump beam power. The inset shows that each cross point is achieved by linear fitting, which is a good  liner approximation when the magnetic field is close to the cross point. (b) The light shift measured by using our approach with respect to different pump beam power. The experimental data is in agreement with a linear fit.}
	\label{fig4}
\end{figure}

Based on the model presented here,
the asymmetric amplitude phenomena can be eliminated with applying a specific magnetic field (namely magic field, $B_{\textbf{magic}}$).
We verify that the magic field has same magnitude and but opposite direction with the light shift of atomic vapor~\cite{Mathur1968}.
This actually suggests a method to measure the light shift of atomic vapor.
When the field is set at the magic field,
the light shift compensates the external magnetic field to effective zero field for the $^{87}$Rb atoms in vapor.
In this condition, the amplitude-frequency response $A_x(\nu, B_{\textbf{magic}})\equiv0$ at any frequencies~\cite{SI},
which results in the symmetric spectra.
We also verify this experimentally.
Figure~\ref{fig4}(a) shows that the magic field follows a linear dependence with the power of the pump beam.
It satisfies well with that the magic field has same magnitude with the light shift of the $^{87}$Rb atomic vapor.
Additionally, we show that the asymmetric ratios of the formic acid doublet are linearly dependent on the power of the pump beam, as shown in Fig.~\ref{fig4}(b).
Similar to the magic field,
there exists a specific pump beam power for observing symmetric spectra.

Asymmetric ultralow-field NMR spectroscopy provides the information of the sign of the Land$\acute{\textrm{e}}$ $g$ factor of $\textrm{NMR}$ systems.
We proof that the regions I and III (see Fig.~\ref{fig3}) have same asymmetry, which is relevant to the sign of the Land$\acute{\textrm{e}}$ $g$ factor of NMR systems.
For example, formic acid doublet both show negative asymmetry in the regions I and $\textrm{III}$.
This implies that the Land$\acute{\textrm{e}}$ $g$ factor of the manifold ($k=1/2, f=1$) in formic acid is positive.
We analyse it as follows.
If the Land$\acute{\textrm{e}}$ $g$ factor is positive, $\nu_1 > \nu_2$ in the region I, and $\nu_1 < \nu_2$ in the region III.
In the region I,
the phase difference at the frequency $\nu_1$ between the magnetometer signals along $x$ and $y$ axes is equal to $-{\pi  \mathord{\left/{\vphantom {\pi  2}} \right.\kern-\nulldelimiterspace} 2} + [{\Phi_x}(\nu_1 ,{B_z}) - {\Phi_y}(\nu_1 ,{B_z})]$,
which is smaller than ${\pi  \mathord{\left/
 {\vphantom {\pi  2}} \right.
 \kern-\nulldelimiterspace} 2}$~\cite{SI}.
While, at the frequency $\nu_2$,
the phase difference, ${\pi  \mathord{\left/{\vphantom {\pi  2}} \right.\kern-\nulldelimiterspace} 2} + [{\Phi_x}(\nu_2 ,{B_z}) - {\Phi_y}(\nu_2 ,{B_z})]$,
is larger than ${\pi  \mathord{\left/
 {\vphantom {\pi  2}} \right.
 \kern-\nulldelimiterspace} 2}$.
Based on the calculation of trigonometric function synthesis (see the Supplemental Material~\cite{SI}), the amplitude of the NMR peak at frequency $\nu_1$ is larger than that at frequency $\nu_2$.
And because of $\nu_1 > \nu_2$, the spectral asymmetry is negative asymmetry.
In the region III, due to $- \pi /2 < {\Phi _x}({\nu _i},{B_z}) - {\Phi _y}({\nu _i},{B_z}) < 0$,
the phase difference of frequency $\nu_1$($\nu_2$) is larger(smaller) than ${\pi  \mathord{\left/
 {\vphantom {\pi  2}} \right.
 \kern-\nulldelimiterspace} 2}$.
Now, the amplitude of the NMR peak at frequency $\nu_1$ is smaller than that frequency $\nu_2$.
In the region III, $\nu_1 >\nu_2$, and the spectra still show negative asymmetry.
Thus, the regions I and III have same negative asymmetry, which corresponds to the positive Land$\acute{\textrm{e}}$ $g$ factor.
Similarly,
when the sign of the Land$\acute{\textrm{e}}$ $g$ factor of relevant manifold is negative,
the NMR spectra show positive asymmetry.
The above analysis is not limited to formic acid, and can be generalized to general NMR systems.

\textsl{Conclusions.\textbf{--}}We have developed a new comprehensive model for explaining the origin of ultralow-field $\textrm{NMR}$ asymmetric spectroscopy,
which keeps unclear before.
The key point in the model is to introduce the response function of phase and amplitude in quantum sensor, e.g., atomic magnetometers,
an important subtlety which is essential for explaining asymmetric ultralow-field $\textrm{NMR}$ spectroscopy.
In the meantime, we demonstrate that there exists a specific magnetic field and pump beam power for achieving symmetric NMR spectra.
It is interesting to note the asymmetric spectroscopy provides an extraordinary tool to obtain additional information of systems,
such as the light shift of atomic vapors and the sign of Land$\acute{\textrm{e}}$ $g$ factor of NMR systems.
Although experimentally demonstrated with atomic magnetometers,
our approach could be applied to a variety of systems, such as using nitrogen-vacancy centres in NMR or electron spin resonance detections~\cite{DeVience2015, Kong2018}.

\textsl{Acknowledgment.\textbf{--}}We thank Dieter Suter, Teng Wu, and Rom$\acute{\textrm{a}}$n Picazo Frutos for useful discussions, Jiankun Chen for preparing the setup diagram, and Kang Dai for providing the atomic vapor cells. This work was supported by National Key Research and Development Program of China (Grant No. 2018YFA0306600), National Natural Science Foundation of China (Grants Nos. 11425523, 11375167, 11661161018, 11227901), Anhui Initiative in Quantum Information Technologies (Grant No. AHY050000), National Science Foundation (Grant ECCS 1710558).

\end{document}